# Fast high fidelity hole spin initialization in a single InGaAs quantum dot

T. M. Godden, S. J. Boyle, A. J. Ramsay, A. M. Fox and M. S. Skolnick

Department of Physics and Astronomy, University of Sheffield, Sheffield, S3 7RH, UK

E-mail: t.godden@sheffield.ac.uk

We demonstrate fast initialization of a single hole spin captured in an InGaAs quantum dot with a fidelity F>99% by applying a magnetic field parallel to the growth direction. We show that the fidelity of the hole spin, prepared by ionization of a photo-generated electron-hole pair, is limited by the precession of the exciton spin due to the anisotropic exchange interaction.

A single spin trapped in a semiconductor quantum dot is a potential qubit with long coherence times[1] and the possibility of picosecond optical control[2]. Recently, there has been considerable interest in the use of hole spins as qubits due to the p-type Bloch functions of the valence band resulting in a suppressed contact hyperfine interaction[3,4] which is the main source of dephasing for electron spins. The initialization of a qubit is a key ingredient of any quantum information processing protocol. Successful approaches of single spin initialization in quantum dots include optical pumping[5,6,7], coherent population trapping[1,8] and the ionization of an electron-hole pair[9,10,11]. Although fidelities F>99.8% have been reported by optical pumping[9], there have been no reports of such fidelities with preparation times comparable to the picosecond gate times used in coherent control experiments.

Previously, we demonstrated a scheme for the fast initialization of a single hole spin (with F=81%), by the ionization of a spin-polarized electron-hole pair[12]. In this letter, we study the dependence of the fidelity on applied magnetic and electric fields. We show that by applying a magnetic field in the growth direction (Faraday geometry), we can achieve near unit fidelity of hole spin preparation, by suppressing the spin mixing generated by the neutral exciton fine structure splitting. We also find that an increased electric-field at B=0 also improves the fidelity by reducing the time available for this spin mixing.

The sample was mounted in a helium bath magneto-cryostat (T=4.2K, B≤5T) and consists of a single layer of InGaAs self assembled quantum dots embedded in the intrinsic region of an n-i-Schottky diode. Details of the layer structure of the wafer can be found in ref. 13. Importantly, in the reverse bias regime, the electron tunnelling rate $\Gamma_e$~30 ps$^{-1}$ ($V_{bias}$=0.8V) is much greater than the rates of hole tunnelling $\Gamma_h$~1 ns$^{-1}$, radiative recombination $\Gamma_r$ ~1 ns$^{-1}$ and the fine structure splitting $\delta_{fs}$~2$\pi$/225 ps$^{-1}$. The slow hole tunnelling rate is due to a hole blocking tunnelling barrier. Therefore if we resonantly excite the neutral exciton transition, the electron quickly tunnels out of the quantum dot, to leave a spin polarized hole.

Before discussing the experimental results, we introduce the principle of operation for the preparation of the single hole spin. Figure 1(a) shows an energy level diagram of the circularly-polarized Zeeman-split neutral exciton states $X^0_\Downarrow = |\Downarrow\uparrow\rangle$ and $X^0_\Uparrow = |\Uparrow\downarrow\rangle$ in a Faraday geometry magnetic field, where $\Uparrow(\Downarrow)$ denotes the hole spin and $\uparrow(\downarrow)$ denotes the electron spin. At time t=0 a circularly polarized laser pulse, termed the preparation pulse, with pulse area $\Theta = \pi$ and FWHM~0.2meV, is resonant with one of the neutral exciton transitions. This creates a spin-polarized electron-hole pair (exciton) which can be detected as a change in photocurrent through the device[14].

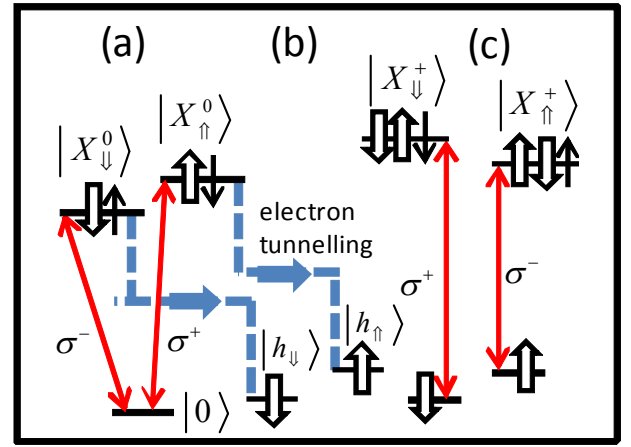

Fig.1. (Color online) Schematic energy level diagram in a magnetic-field of (a) neutral exciton $X^0$ (b) hole spin states $h_{\Downarrow(\Uparrow)}$ (c) Positive trion transitions $X^+$. (a) Preparation pulse: $\sigma^\pm$ polarized $\pi$ pulse resonant with $X^0$. (b) Ionization: electron tunnels out of the dot with rate $\Gamma_e$~30ps$^{-1}$ (dashed line). (c) Detection: $\sigma^\pm$ polarized $\pi$ pulse resonant with $X^+$.

The anisotropic exchange interaction couples the exciton states $X^0_\Downarrow$ and $X^0_\Uparrow$ and causes the exciton spin to precess. This is modelled by a 2x2 Hamiltonian with a Zeeman splitting of $\hbar\omega_z$ and a fine structure splitting $\hbar\delta_{fs}$. The eigenvectors are $|\psi_+\rangle = \sin\theta|\Uparrow\downarrow\rangle + \cos\theta|\Downarrow\uparrow\rangle$ and $|\psi_-\rangle = \cos\theta|\Uparrow\downarrow\rangle - \sin\theta|\Downarrow\uparrow\rangle$ where $\theta$ is a mixing angle given by $\tan(2\theta)$=-$\delta_{fs}/\omega_z$. A $\sigma^-$ polarized preparation pulse creates an exciton of spin $|\Downarrow\uparrow\rangle$, which evolves into $|\psi(t)\rangle = \sin\theta|\psi_+\rangle e^{i\lambda t} + \cos\theta|\psi_-\rangle e^{-i\lambda t}$, where $\lambda = 0.5\sqrt{\omega_z^2 + \delta_{fs}^2}$ is the eigen-energy. Projecting on to the exciton spin bases, we find the exciton spin state populations $P_{\Downarrow\uparrow} = \sin^2(2\theta)\sin^2(\lambda t)$ and $P_{\Uparrow\downarrow} = 1 - P_{\Downarrow\uparrow}$. Hence the difference in the neutral exciton spin populations is $P_{\Uparrow\downarrow} - P_{\Downarrow\uparrow} = (1 - 2\sin^2(2\theta)\sin^2(\lambda t))e^{-\Gamma_x t}$ and the total neutral exciton population is $P_{\Uparrow\downarrow} + P_{\Downarrow\uparrow} = e^{-\Gamma_x t}$, where the phenomenological exciton decay rate $\Gamma_x = \Gamma_e + \Gamma_r \approx \Gamma_e$ is introduced.

Fig 1.(b) shows how the exciton spin states are mapped onto the hole spin states after fast ionization under the applied electric-field. Neglecting hole spin relaxation[15], the rate equations for the Zeeman-split hole spin state populations ($h_\Downarrow$ and $h_\Uparrow$) are $\dot{h}_\Downarrow = \Gamma_e P_{\Downarrow\uparrow} - \Gamma_h h_\Downarrow$ and $\dot{h}_\Uparrow = \Gamma_e P_{\Uparrow\downarrow} - \Gamma_h h_\Uparrow$.

Using the expressions for the exciton state populations and integrating over time, we find the hole spin state contrast $C$ and fidelity $F$:

$$C = 2F - 1 = \lim_{\Gamma_x t \gg 1}\left(\frac{h_\Downarrow - h_\Uparrow}{h_\Downarrow + h_\Uparrow}\right) = 1 - \left(\frac{\delta_{fs}^2}{\delta_{fs}^2 + \omega_z^2 + (\Gamma_X - \Gamma_h)^2}\right) \quad \text{Eq.1}$$

The fidelity $F$ in equation 1 is a measure of the purity of the preparation of a single hole spin. From Eq.1 we conclude that (i) $F$ is limited by a competition between the fine-structure splitting and electron tunnelling rate, and (ii) For large B-fields where $\omega_z^2 \gg \delta_{fs}^2$, $F \to 1$ since the eigenstates are transformed from linear to circular.

In order to determine the frequency to be employed for the preparation pulse, we excite the dot with a single laser pulse, and measure the photocurrent as the frequency of the laser is scanned through the neutral exciton resonances. Figs. 2a (B=0T) and 2c (B=3T) show the single pulse preparation of the neutral states $X_\Downarrow^0$ ($X_\Uparrow^0$) using $\sigma^-$ ($\sigma^+$) polarization with a gate voltage of 0.8V. Under an applied magnetic field (Fig.2c), the absorption peaks are split by the exciton Zeeman energy $\hbar\omega_z = g\mu_B B$. An exciton g-factor $g=1.69\pm0.05$ was measured for both the neutral and positive exciton transitions.

To measure the purity of the hole spin preparation, we use a second circularly-polarized time-delayed pulse, termed the detection pulse, with a pulse area $\Theta=\pi$, and FWHM of ~0.2meV. With a $\sigma^+$ preparation pulse resonant with the $0 - X_\Uparrow^0$ transition, we monitor the change in photocurrent as the detection pulse is scanned through the $h - X^+$ transition. For all measurements, a time delay between preparation and detection of $\Delta t = 160$ps is chosen to maximise the $X^+$ signal at 0.8V. The selection rules of the positive trion transition are presented in the energy-level diagram of fig. 1(c). Absorption of the detection pulse as it is scanned through resonance with the $h - X^+$ transitions is conditional on the spin of the hole. In the case of perfect spin preparation, absorption of a detection pulse co-polarized with respect to the preparation pulse is forbidden due to Pauli blocking. By contrast, absorption of a cross-polarized detection pulse is allowed, resulting in a change of photocurrent proportional to the occupation of the hole spin state, selected by the polarization of the detection pulse.

Figures 2(b,e) show the photocurrent absorption spectra for two pulse measurements of $X_\Uparrow^+$ and $X_\Downarrow^+$ at B=0T and B=3T respectively. In figs. 2(b,e), a photocurrent peak corresponding to the $h_\Uparrow - X_\Uparrow^+$ transition is clearly observed for cross-polarized excitation, with a peak change in photocurrent of $PC_{+-}$. In fig. 2(b), in the case of zero applied magnetic field, a smaller trion peak of amplitude $PC_{++}$ is also observed for co-polarized excitation. This indicates that the hole has not been prepared purely in the $h_\Uparrow$ state. Under an applied magnetic field, as presented in fig. 2(e), the trion peak cannot be observed in the case of co-polarized excitation, indicating a high purity hole spin state has been prepared. Fig.2(e) shows the two pulse measurement with B=-3T. In this case the $X_\Downarrow^0$ state is prepared. Again no $X^+$ peak is observed for co polarized pulses, confirming the high fidelity spin preparation.

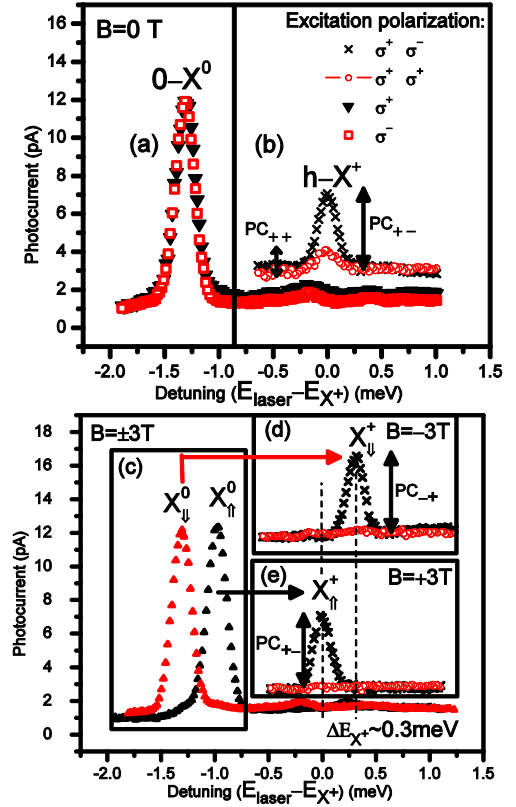

Fig.2. (Color online) Photocurrent spectra. (a) and (c) single pulse spectra of neutral exciton using $\sigma^\pm$ excitation with B=0T and |B|=3T. (b), (d) and (e) Two pulse spectra of $X^+$ using cross (crosses) and co (circles) $\sigma^\pm$ polarization with |B|=3T. The absence of the $X^+$ peak for co-circular excitation indicates high purity spin preparation has been achieved.

The purity of hole spin preparation is quantified by the experimental fidelity $F_{PC}=PC_{-+}/(PC_{++}+PC_{+-})$. Figure 3(a) shows the measured fidelity as a function of applied magnetic field at a gate voltage of 0.8V. The fidelity is observed to increase strongly as a function of B from 81% at B=0T to ~100% for B>1T. We fit the experimental data to Eq.1 using two fitting parameters $\delta_{fs}=2\pi/(225\pm25)$ps$^{-1}$ and $(\Gamma_X-\Gamma_h)=1/(28\pm4)$ps$^{-1}$. This is in good agreement with previous measurements of $\Gamma_X$ and $\delta_{fs}$ on this dot[13,16].

The dependence of the fidelity on the electron tunnelling rate was also investigated by measuring $F_{PC}$ as a function of gate-voltage at B=0T. The results are presented in fig. 3(b). For increasing gate-voltage and hence electron tunnelling rate, we observe a rise in the measured fidelity. The red lines show a calculation of the range of possible values for $F$ using known tunnelling rates[13] and the fine structure splitting determined from the fitting in fig.3(a). We find good agreement between the model and the experiment. A contributing factor to any discrepancy may be related to neglecting a small variation in the fine-structure splitting with gate-voltage[17].

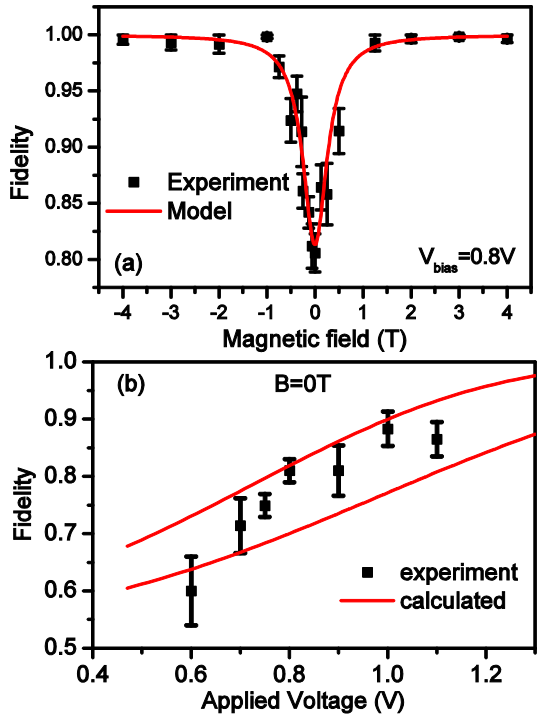

Fig. 3 (Color online) **(a)** Hole spin preparation fidelity as a function of magnetic field, the red line is a fit to Eq.1. (b) Hole spin preparation fidelity as a function of applied reverse bias voltage at B=0T. Two (red) lines indicate range of calculated values of *F* using Eq.1

The results in fig.3(b) indicate that increasing the gate voltage improves the spin preparation by decreasing the time available for the exciton spin to precess. However, by increasing the static gate voltage there is a trade off between improving the fidelity of the hole spin preparation and the lifetime, and hence coherence time of the hole. In principle, this may be overcome through dynamic control of the tunnelling rates using a voltage modulation[18]. Further improvements could be made by optimizing the tunnel barriers in the device to achieve faster and slower electron and hole tunnelling rates respectively.

To summarize, we demonstrate the fast, (1/e time of ~30 ps) triggered, high fidelity (F>99%) initialization of a single hole spin in a Faraday geometry magnetic field using a ps laser pulse and an electrical detection technique. The purity of the hole spin preparation is limited by the mixing between the electron and hole spins, generated by the anisotropic exchange interaction before the electron tunnels from the dot. For the dot presented here, with a fine-structure splitting of $\delta_{fs}=2\pi/(225\pm25)$ps$^{-1}$, a gate voltage of 0.8 V, and a strong 4-T magnetic field, our model predicts a fidelity of 99.9% for the hole spin preparation. By reducing $\delta_{fs}$ and the strength of spin mixing, for example by using techniques such as thermal annealing[19], similarly high fidelities could be achieved at significantly lower magnetic fields.

The authors acknowledge financial support from EPSRC UK EP/G001642 and the QIPIRC UK.